\RequirePackage{lineno}
\documentclass[ALICE,manyauthors]{cernphprep}
\usepackage{graphicx}
\usepackage{xspace}

\providecommand*\ie{i.e.\xspace}

\providecommand*\sigmabb{\sigma_{{\rm b}\bar{{\rm b}}}\xspace}

\providecommand*\gevc{GeV/$c$\xspace}

\providecommand*\ppbar{p$\bar{\rm p}$\xspace}
\providecommand*\pbpb{Pb-Pb\xspace}

\providecommand*\ccbar{{\rm c}\bar{{\rm c}}\xspace}
\providecommand*\bbbar{{\rm b}\bar{{\rm b}}\xspace}

\providecommand{\jpsi}{${\rm J}/\psi$\xspace}

\providecommand{\pt}{\ensuremath{p_{\rm T}}\xspace}

\providecommand{\dedx}{d$E$/d$x$\xspace}

\providecommand*{\pp}{pp\xspace}
\providecommand{\ppbar}{p\bar{p}\xspace}
\providecommand{\dzero}{\ensuremath{d_{0}}\xspace}

\begin{document}
\begin{titlepage}
\PHnumber{2012-229}
\PHdate{08 August 2012}

\title{Measurement of electrons from beauty hadron decays\\in pp collisions 
       at $\sqrt{s}=7$ TeV}
\Collaboration{ALICE Collaboration%
         \thanks{See Appendix~\ref{authorlist} for the list of collaboration
                      members}}
\ShortAuthor{ALICE Collaboration}
\ShortTitle{Electrons from beauty hadron decays in pp collisions at
            $\sqrt{s} = 7$ TeV}
\begin{abstract}
The production cross section of electrons from semileptonic decays of 
beauty hadrons was measured at mid-rapidity ($|y| < 0.8$) in the 
transverse momentum range $1 < \pt < 8$~\gevc with the ALICE experiment 
at the CERN LHC in \pp collisions at a center of mass energy $\sqrt{s} = 7$~TeV
using an integrated luminosity of 2.2~nb$^{-1}$. Electrons from beauty hadron 
decays were selected based on the displacement of the decay vertex from the 
collision vertex. A perturbative QCD calculation agrees with the measurement 
within uncertainties. The data were extrapolated to the full phase space to 
determine the total cross section for the production of beauty quark-antiquark 
pairs.                                                   
\end{abstract}
\end{titlepage}
\setcounter{page}{2}
\maketitle

The measurement of heavy-flavor (charm and beauty) production in proton--proton
(\pp) collisions at the CERN Large Hadron Collider (LHC) provides a crucial
testing ground for quantum chromodynamics (QCD), the theory of strong 
interactions, in a new high-energy regime. Because of their large masses heavy 
quarks are mainly produced via initial hard parton-parton collisions,
even at low transverse momenta \pt. Therefore, heavy-flavor production 
cross sections constitute a prime benchmark for perturbative QCD (pQCD) 
calculations. Furthermore, heavy-flavor measurements in \pp collisions provide 
a mandatory baseline for corresponding studies in nucleus-nucleus collisions. 
Heavy quark observables are sensitive to the properties of the strongly 
interacting partonic medium which is produced in such collisions.

Earlier measurements of beauty production in \ppbar collisions at 
$\sqrt{s} = 1.96$~TeV at the Tevatron~\cite{cdf_beauty} are in good 
agreement with pQCD calculations at fixed order with next-to-leading log 
resummation (FONLL)~\cite{fonll1,fonll2}. Measurements of charm production,
available at high \pt only~\cite{cdf_charm}, are close to the upper limit 
but still consistent with such pQCD calculations. The same trend was observed 
in \pp collisions at $\sqrt{s} = 0.2$~TeV at RHIC~\cite{phenix_e,star_d}.

In \pp collisions at the LHC, heavy-flavor production was investigated 
extensively at $\sqrt{s} = 7$~TeV in various decay channels. With LHCb 
beauty hadron production cross sections were measured at forward 
rapidity~\cite{lhcb_b} and, at high \pt only, with CMS at 
mid-rapidity~\cite{cms_bszero}. At low \pt, mid-rapidity 
\jpsi meson production from beauty hadron decays was studied with
ALICE~\cite{alice_jpsi_b}. These results, as well as the 
mid-rapidity D-meson production cross sections measured with 
ALICE~\cite{alice_d2h}, are well described by FONLL pQCD 
calculations. The same is true for the production cross sections of electrons 
and muons from semileptonic decays of heavy-flavor hadrons reported by 
ATLAS~\cite{atlas_lepton} at high \pt, and by ALICE down to low 
\pt~\cite{alice_e,alice_muon}. However, still missing at the LHC is the separation of 
leptons from charm and beauty hadron decays at low \pt, which is important 
for the total beauty production cross section and which provides a crucial 
baseline for \pbpb collisions.

This Letter reports the mid-rapidity ($|y| < 0.8$) production cross section 
of electrons, $({\rm e}^+ + {\rm e}^-)/2$, from semileptonic beauty hadron 
decays measured with the ALICE experiment in the range $1 < \pt < 8$~\gevc in 
\pp collisions at $\sqrt{s} = 7$~TeV. Two independent techniques were used for 
the separation of beauty hadron decay electrons from those originating 
from other sources, in particular charm hadron decays. The resulting invariant 
cross sections of electrons from beauty and from charm hadron decays
are compared with corresponding predictions from a FONLL pQCD calculation. In
addition, the measured cross sections were extrapolated to the full phase space
and the total beauty and charm production cross sections were determined.

The data set used for this analysis was recorded during the 2010 LHC run with 
ALICE, which is described in detail in~\cite{alice}. 
Charged particle tracks were reconstructed in the pseudorapidity range 
$|\eta| < 0.8$ with the Time Projection Chamber (TPC) and the Inner Tracking 
System (ITS) which, in addition, provides excellent track spatial resolution 
at the interaction point. Electron candidates were selected with the TPC and the 
Time-Of-Flight detector (TOF). Data were collected using a minimum bias (MB) 
trigger~\cite{alice_e} derived from the VZERO scintillator arrays and the 
Silicon Pixel Detector (SPD), which is the innermost part of the ITS consisting
of two cylindrical layers of hybrid silicon pixel assemblies. The MB trigger 
cross section $\sigma_{\rm MB} = 62.2 \pm 2.2$~mb~\cite{alice_cross_section} 
was measured in a van-der-Meer scan. An integrated luminosity of 2.2~nb$^{-1}$ 
was used for this analysis.

Pile-up events were identified by requiring no more than one primary vertex 
to be reconstructed with the SPD as discussed in~\cite{alice_e}. Taking into 
account the efficiency of the pile-up event identification, only 2.5\% of the triggered 
events suffered from pile-up. The corresponding events were removed from 
the analyzed data sample. The systematic uncertainty due to the remaining 
undetected pile-up events was negligible.

Events and tracks were selected following the approach from a previous 
analysis~\cite{alice_e}.
Charged particle tracks reconstructed in the TPC and ITS were propagated 
towards the outer detectors using a Kalman filter approach~\cite{TrackingKF}. 
Geometrical matching was applied to associate tracks with hits in the outer detectors. 
To guarantee good particle identification based on the specific \dedx
in the TPC, tracks were required to include a minimum number of 80 clusters
used for the energy loss calculation. A cut on the number of clusters for
tracking is used to enhance the electron/pion separation. The stringent request 
for at least 120 clusters from the maximum of 159 enhances electrons relative to hadrons. 
In total, at least four ITS hits were required to be associated with a track.
A cut on the distance of closest approach (DCA) to the primary vertex 
in the plane perpendicular to the beam axis ($xy$) as well as in the beam 
direction ($z$) was applied to reject background tracks and non-primary tracks. 
Differently from the heavy-flavor electron analysis~\cite{alice_e}, the pseudorapidity range was extended 
to $|\eta| < 0.8$, and tracks were required to be associated with hits in both layers 
of the SPD in order to minimize the contribution from tracks with randomly associated 
hits in the first pixel layer. The latter criterion provides a better measurement 
of the track's transverse impact parameter \dzero, \ie the DCA to the primary collision 
vertex in the plane perpendicular to the beam axis, where the sign of \dzero is 
attributed on the basis of the relative position of primary vertex and 
the track prolongation in the direction perpendicular to the direction 
of the transverse momentum vector of the track.

Electron candidates were required to be consistent within three standard
deviations with the electron time of flight hypothesis, thus efficiently 
rejecting charged kaon background up to momenta of $\approx 1.5$~\gevc and
proton background up to $\approx 3$~\gevc. Additional background, in particular
from charged pions, was rejected using the specific energy loss, \dedx, 
measured for charged particles in the TPC. 

\begin{figure}[t]
\centering
\includegraphics[width=0.7\linewidth]{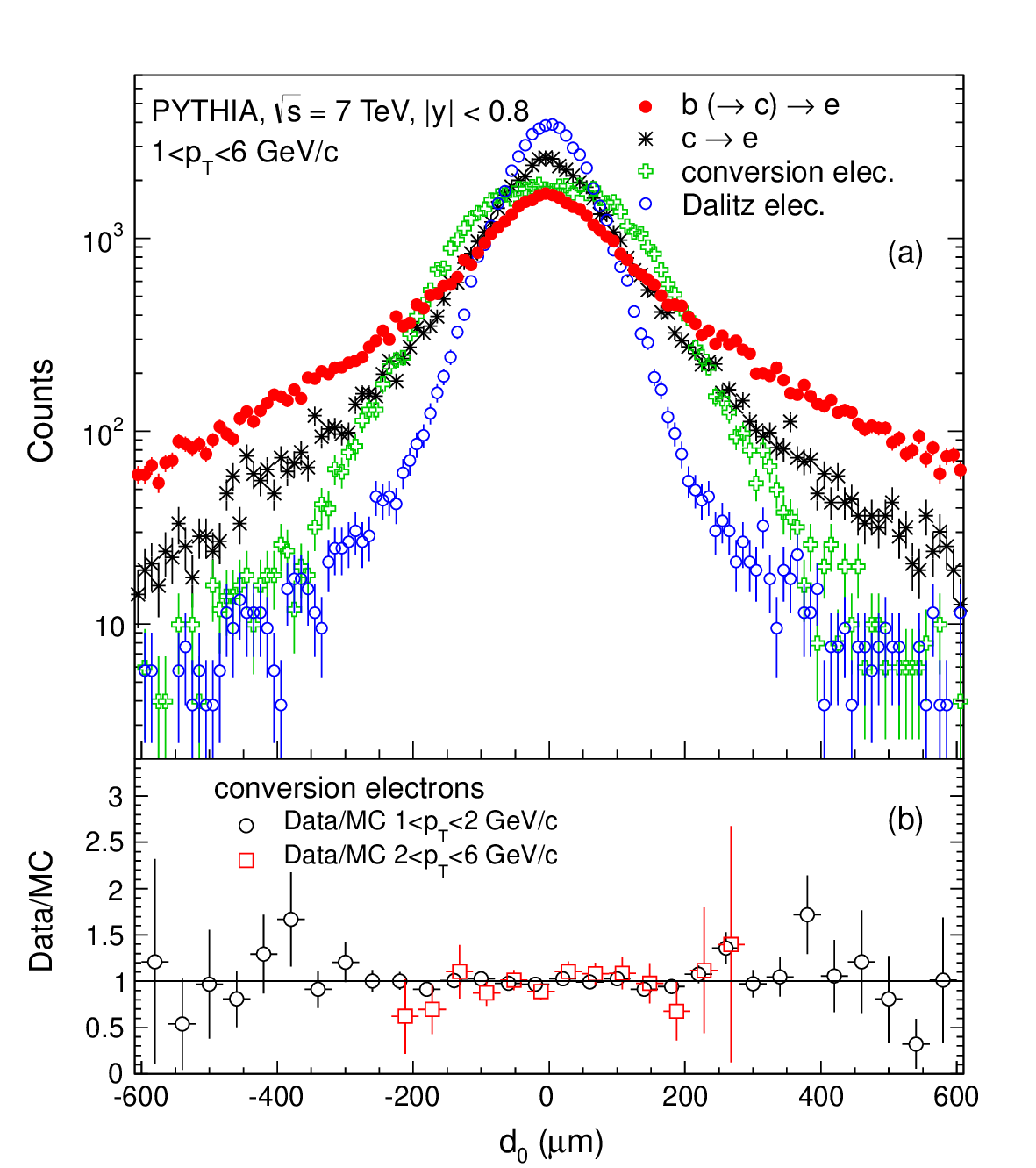}
\caption{\label{fig:fig1}
         (Color online) (a) \dzero distributions of electrons from beauty and 
         charm hadron decays as well as from decays of light hadrons and from 
         photon conversions obtained from PYTHIA simulations in the electron 
         \pt range $1 < \pt < 6$~\gevc. The distributions were normalized to
         the same integrated yield. (b) Ratios of the measured and the 
         simulated \dzero distributions of conversion electrons in the ranges 
         $1 < \pt < 2$~\gevc and $2 < \pt < 6$~\gevc (points shifted in \dzero 
         by 10 $\mu$m for better visibility).}   
\end{figure}

Due to their long lifetime ($c\tau \sim 500$~$\mu$m), 
beauty hadrons decay at a secondary vertex displaced in space from the primary 
collision vertex. Consequently, electron tracks from semileptonic beauty 
hadron decays feature a rather broad \dzero distribution, as indicated 
by simulation studies in Fig.~\ref{fig:fig1}(a). Also shown are the 
\dzero distributions of the main background sources, 
\ie electrons from charm hadron decays, from Dalitz and dilepton decays of 
light mesons, and from photon conversions. These distributions were obtained
from a detailed Monte Carlo simulation of the experiment using 
GEANT3~\cite{geant}. With the PYTHIA 6.4.21 event generator~\cite{pythia} 
\pp collisions were produced employing the Perugia-0 parameter 
tuning~\cite{perugia}. 
The \pt shapes of beauty hadron decay electrons from a FONLL pQCD calculation~\cite{fonll3} 
and from PYTHIA are in good agreement. The PYTHIA simulation does not 
reproduce precisely the \pt-differential yields of background sources measured in data.
Therefore, the \pt distributions of the relevant electron 
sources in PYTHIA were re-weighted to match the distributions measured with ALICE,
prior of propagation through the ALICE apparatus using GEANT3.
After the full Monte Carlo simulation, the same event cuts and track selection criteria 
(including that on \dzero) as in data were applied. The \pt distributions of the backgrounds 
were normalized by the number of events passing these event selection cuts, 
corrected for the efficiency to reconstruct a primary vertex. Background electrons surviving 
these selection criteria were subtracted from the inclusive electron spectrum obtained from data. 
This approach relies on the availability of the \pt-differential cross section measurements
of the main background sources.

The production cross sections of $\pi^0$ and $\eta$ mesons, the 
dominant sources of electrons from Dalitz decays and from photons which convert 
in material into ${\rm e}^+{\rm e}^-$ pairs, were measured with ALICE 
in \pp collisions at \mbox{$\sqrt{s} = 7$~TeV~\cite{alice_pieta}}. 
The conversion electron yield depends on the material budget which was
measured with a systematic uncertainty of 4.5\%~\cite{alice_pieta}.
Other light hadrons and heavy quarkonia contribute through their decays to the 
electron spectrum and their phase space distributions were calculated 
with the approach described in ~\cite{alice_e}. This calculation also includes real and 
virtual photon production via partonic hard scattering processes.
D$^0$, D$^+$, and D$_{\rm s}^+$ meson production cross sections were measured
with ALICE~\cite{alice_d2h,alice_ds} in the transverse momentum ranges 
$1 < \pt < 16$~\gevc, $1 < \pt < 24$~\gevc, and $2 < \pt < 12$~\gevc, 
respectively. Based on a FONLL pQCD calculation~\cite{fonll3} the measured 
\pt-differential cross sections were extrapolated to $\pt = 50$~\gevc. 
The contribution from the unmeasured high-\pt region to the electron yield 
from D-meson decays was estimated to be $\leq 10$\% for electrons with 
$\pt < 8$~\gevc. A contribution from $\Lambda_c$ decays was included using a 
measurement of the ratio $\sigma(\Lambda_c)/\sigma({\rm D}^0 + {\rm D}^+)$ from
ZEUS~\cite{zeus_lambdac}.

The measured \pt spectra of the main background sources drop more quickly 
with \pt than the ones generated by PYTHIA for \pt $>$ 1~\gevc.
The ratio of the measured yield and the yield from PYTHIA, which was 
used to weight the spectra of the electron sources in PYTHIA, is 
1.3 (0.6) at $\pt = 1 \, (10)$~\gevc for $\pi^0$.
The corresponding ratio is 2.4 (1.3) at $\pt = 1 \, (10)$~\gevc for $\eta$ mesons,
and 0.95 (0.2) at $\pt = 1 \, (10)$~\gevc for electrons from charm hadron decays. 

A cut on the \dzero parameter is applied in order to enhance 
the signal-to-background ratio (S/B) of electrons from beauty hadron decays.
For this, it is crucial that the \dzero resolution is properly reproduced in the simulation. 
The \dzero resolution is found to be 80~$\mu$m (30~$\mu$m) 
for tracks with $\pt = 1 \, (10)$~\gevc~\cite{alice_d2h}. 
The agreement of the \dzero measurement of electron candidates with 
the simulation is demonstrated in Fig.~\ref{fig:fig1}(b), which shows the ratios of the measured \dzero 
distribution to the one from simulation in the \pt ranges $1 < \pt < 2$~\gevc 
and $2 < \pt < 6$~\gevc for electrons from photon conversions, which is the only identifiable source in data. 
A pure sample of electrons from photon conversions in the detector material 
was identified using a V0-finder and topological cuts~\cite{V0}.
At \mbox{\pt $> 6$~\gevc}, the number of reconstructed conversions was 
statistically insufficient for this cross check. 
In addition, the \dzero resolution measured for charged tracks 
in data is reproduced within 10 \% by the Monte Carlo simulation~\cite{alice_d2h}.
The difference in the particle multiplicities between data and simulation
gives an effect on the primary vertex resolution, which is included 
in the \dzero resolution as a convolution of the track position and the primary vertex resolution. 
The Monte Carlo simulation shows that the electron Bremsstrahlung effect 
is limited to transverse momenta below 1 \gevc. 
At higher \pt, the particle species dependences of the \dzero resolution is negligible.

\begin{figure}[t]
\centering
\includegraphics[width=0.49\linewidth]{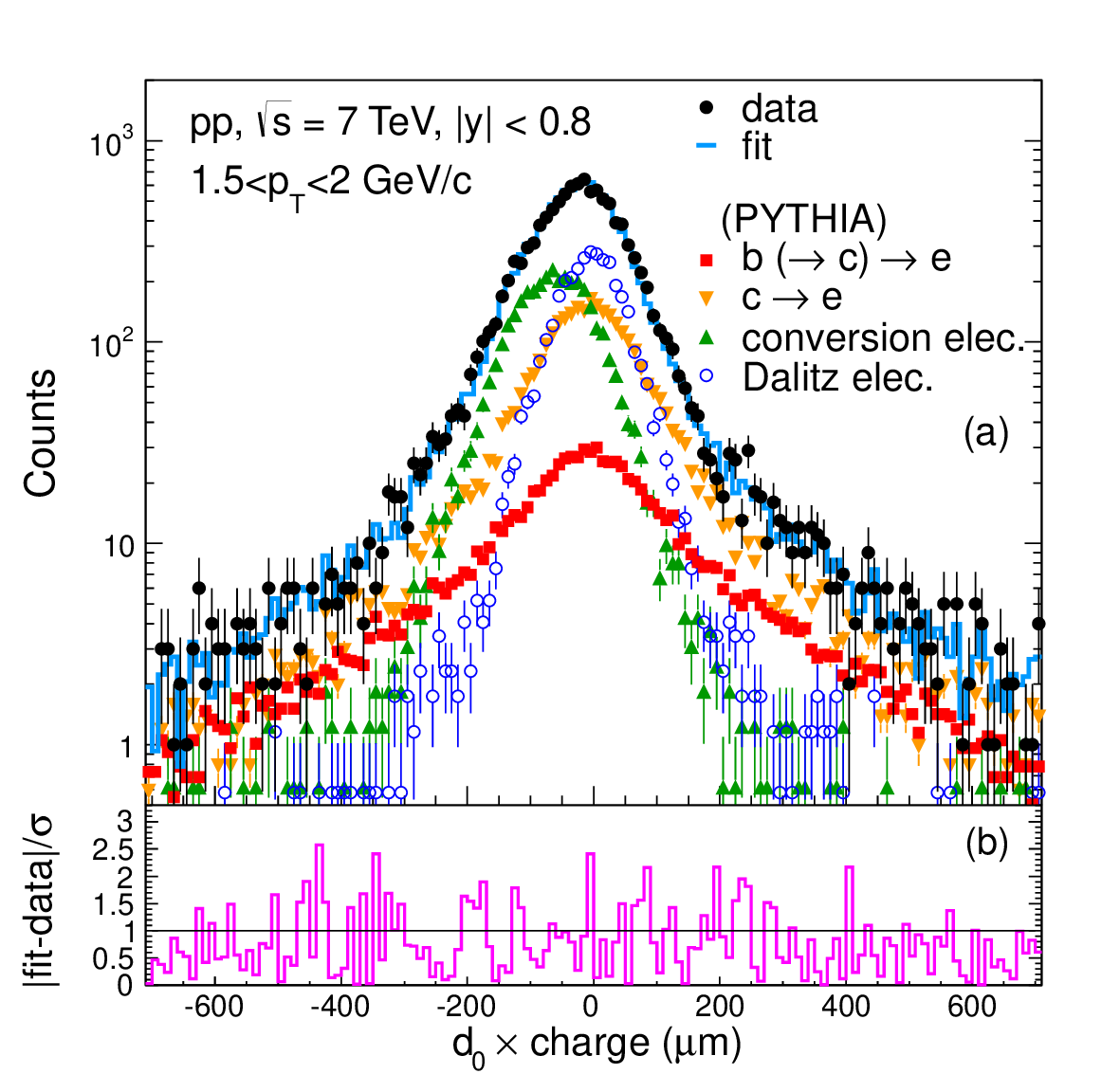}
\includegraphics[width=0.49\linewidth]{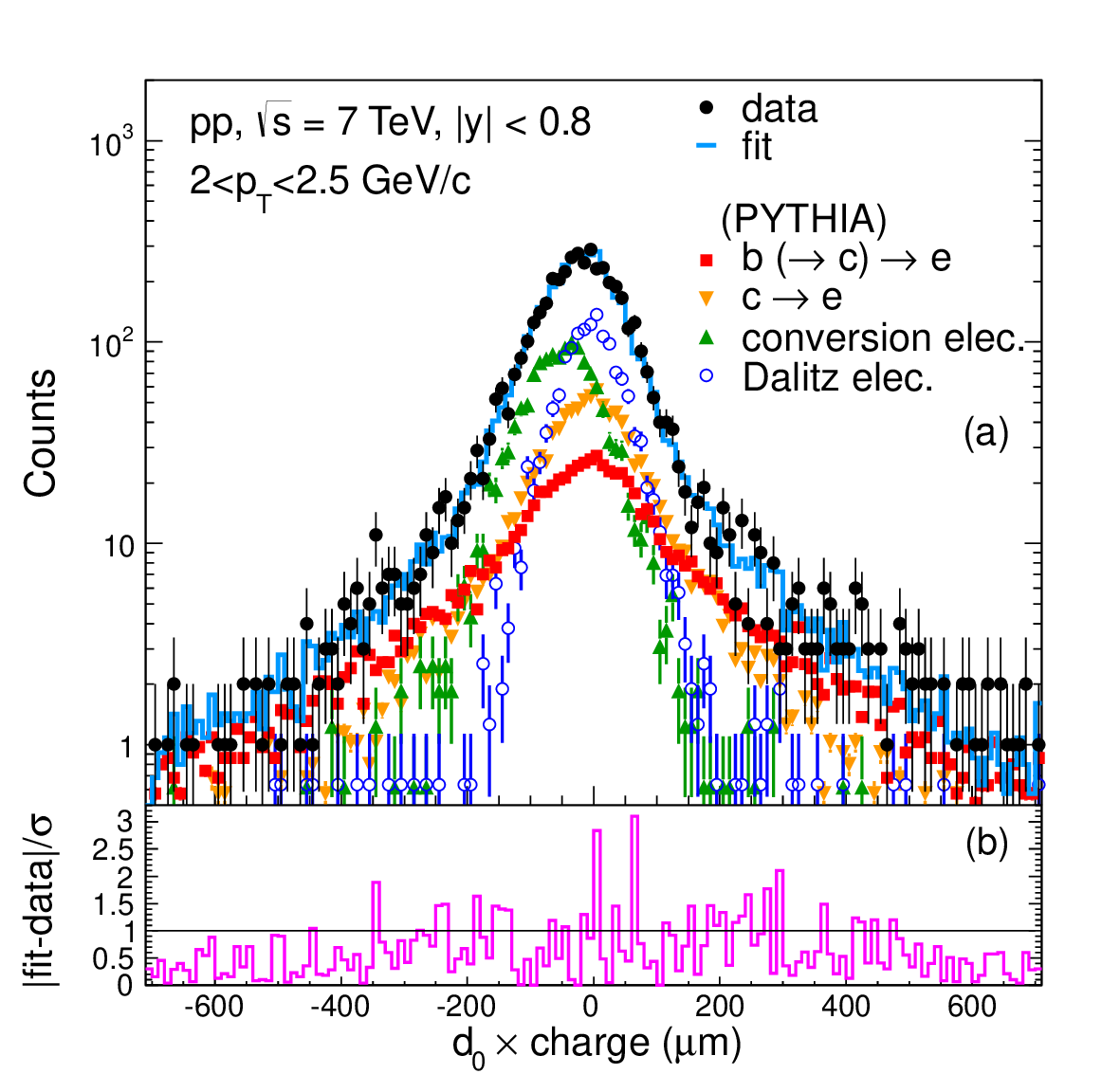}
\caption{\label{fig:fig2}
          (Color online) (a) Distribution of \dzero$\times$ charge for  electron candidates 
          after all analysis cuts (except that on \dzero) superimposed to the best-fit result. 
          The fit function is defined as the sum of the Monte Carlo  \dzero distribution of 
          beauty electrons and those of electrons from all other sources, the normalizations 
          being the free parameters in the fit. The error bars represent the statistical uncertainties. 
          (b) Differences between the data and the best fit result divided by the statistical error.} 
\end{figure}

Figure~\ref{fig:fig2} shows that the d$_{\rm 0}$ distribution of the data sample
is well described by the cocktail of signal and background. The measured \dzero 
distribution of identified electrons was fitted by minimizing a $\chi^{2}$ between 
the measured \dzero distribution and the sum of the Monte Carlo \dzero distributions 
of signal and background in the corresponding electron \pt range. The differences 
between the data and the cocktail are consistent with statistical variations.
The ratio of the signal to background yields, which is obtained by this fit procedure, 
agrees with that obtained in the present analysis within statistical uncertainties. 

The widths of the \dzero distributions depend on \pt. Only electrons satisfying the condition
$|d_0| > 64 + 780 \times \exp(-0.56 \,\pt)$ (with $d_0$ in $\mu$m and
\pt in \gevc) were considered for the further analysis. This \pt-dependent \dzero cut was determined 
from the simulation to maximize the significance for the beauty decay electron spectrum.
The possible bias introduced by this optimization is taken into account in the estimation 
of the systematic uncertainties, by varying substantially the cut value.

\begin{figure}[t]
\centering
\includegraphics[width=0.7\linewidth]{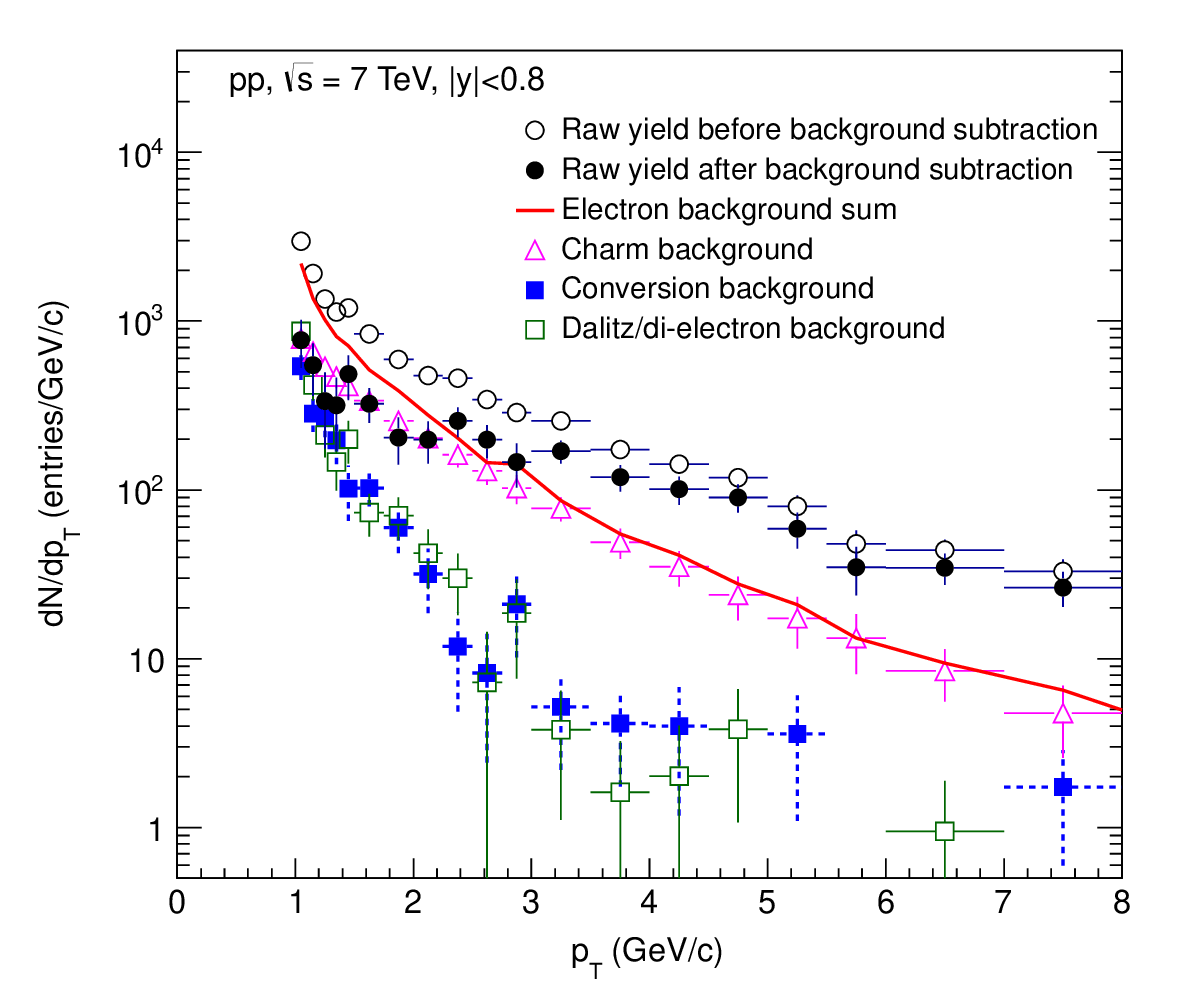}
\caption{\label{fig:fig3}
         (Color online) The signal (black solid circle) and the background yields after the application 
         of the track selection criteria including the one on \dzero. The background electrons 
         (red solid line), \ie the sum of the electrons from charm hadron decays, from Dalitz and 
         dilepton decays of light mesons, and from photon conversions, were subtracted from 
         the inclusive electron spectrum (black open circle). The error bars represent the statistical 
         uncertainties. The symbols are plotted at the center of each bin.}
\end{figure}

Fits of the TPC \dedx distribution in momentum slices indicate that the 
remaining hadron contamination grows from less than 10$^{-5}$ at 1~\gevc to 
$\approx 20$\% at 8~\gevc before the application of the \dzero cut.
Since hadrons originate from the primary collision vertex, the latter cut
reduces the remaining hadron contamination to less than 3\% even at the
highest \pt considered here. The electron background from sources other than
beauty hadron decays was estimated based on the method described above. 
In Figure~\ref{fig:fig3} the raw electron yield, as well as the non-beauty electron 
background yield, which is subtracted in the analysis, are shown 
after the application of the track selection criteria.
At $\pt = 1$~\gevc, the background contributions from charm hadron
decays, light meson decays, and photon conversions are approximately equal
and S/B is $\approx 1/3$. At $\pt = 8$~\gevc, the background originates 
mostly from charm hadron decays and S/B is $\approx 5$. 

The electron yield from beauty hadron decays, $N_e(\pt)$, was corrected 
for the geometrical acceptance, the track reconstruction efficiency, the 
electron identification efficiency, and the efficiency of the \dzero cut. 
The total efficiency $\varepsilon$ is the product of these individual factors.
$\varepsilon$ was computed from a full detector simulation using GEANT3 
as discussed in~\cite{alice_e}.
In addition, the electron \pt distribution was corrected for effects of finite 
momentum resolution and energy loss due to Bremsstrahlung via a \pt unfolding 
procedure which does not depend on the \pt shape of  Monte Carlo 
simulation~\cite{alice_e}. 

The invariant cross section of electron production from beauty hadron decays 
in the range $|y| < 0.8$ was then calculated using the corrected electron \pt 
spectrum, the number of minimum bias \pp collisions $N_{\rm MB}$, and the 
minimum bias cross section $\sigma_{\rm MB}$ as

\begin{equation}
\frac{1}{2\pi\pt} \frac{d^{2}\sigma}{d\pt dy} = 
\frac{1}{2\pi\pt^{\rm{c}}} 
\frac{N_e(\pt)}{\Delta y \Delta \pt}
\frac{1}{\varepsilon}
\frac{\sigma_{\rm MB}}{N_{\rm MB}},
\label{eq:cross_section}
\end{equation}
where $\pt^{\rm{c}}$ are the centers of the \pt bins with widths $\Delta \pt$
and $\Delta y = 0.8$ is the width of the rapidity interval.

A summary of the estimated relative systematic uncertainties is provided in Table~\ref{sys_overview}.
The systematic uncertainties for the tracking and the particle identification are the following:
the corrections of the ITS, TPC, TOF tracking efficiencies, 
the TOF, TPC particle identification efficiencies, 
the \pt unfolding procedure.
These amount to $^{+17}_{-14} (^{+8}_{-14})$\% for \pt $<$($>$) 3~\gevc. 
Additional systematic uncertainties specific for this analysis due to 
the \dzero cut, the subtraction of the light hadron decay background 
and charm hadron decay background were added in quadrature.
The systematic uncertainty induced by the \dzero cut was evaluated 
by repeating the full analysis with modified cuts. The variation of this cut 
was chosen such that it corresponds to $\pm 1 \sigma$, where $\sigma$ 
is the \dzero resolution measured on data~\cite{alice_d2h}.
These vary the minimum \dzero cut efficiency by $\pm 20$\%.
In addition, the full analysis was repeated after smearing 
the \dzero resolution in the Monte Carlo simulation by 10\%~\cite{alice_d2h},
considering the maximum differences in the \dzero distribution in data and simulation. 
The uncertainty due to the background subtraction was evaluated by propagating 
the statistical and systematic uncertainties of the light and charm hadron 
measurements used as analysis input.
At low \pt, the uncertainties are dominated by the subtraction of charm 
hadron decay background.

\begin{table}
\begin{center}
\caption{Overview of the contributions to the systematic uncertainties. 
The total systematic uncertainty is calculated as the quadratic sum of all contributions.}
\begin{tabular}[ht]{l|c}
\hline\hline
\pt range (\gevc)& 1 -- 8\\
\hline
\hline
Error source & \multicolumn{1}{c}{systematic uncertainty [\%]}\\
\hline
Track matching & $\pm$2\\
ITS number of hits & $^{+1}_{-4}$\\
TPC number of tracking clusters & $^{+15}_{-7}$($^{+3}_{-4}$) for \mbox{\pt $<$ 2.5($>$2.5)~\gevc}\\
TPC number of PID clusters & $\pm$2\\
DCA to primary vertex in xy (z) & $\pm$1\\
TOF matching and PID & $\pm$5\\
TPC PID & $+5$($^{+2}_{-5}$) for \mbox{\pt $<$ 3($>$3)~\gevc}\\
Minimum \dzero cut & $\pm$12\\
Charge dependence & $^{+1}_{-7}$\\
$\eta$ dependence & $-6$\\
Unfolding & $\pm$5\\
Light hadron decay background & $\approx$10($<$2) for \mbox{\pt = 1($>$2)~\gevc}\\
Charm hadron decay background & $\approx$30($<$10) for \pt = 1($>$3)~\gevc\\
\hline\hline
\end{tabular}
\label{sys_overview}
\end{center}
\end{table}

\begin{figure}[t]
\centering
\includegraphics[width=0.7\linewidth]{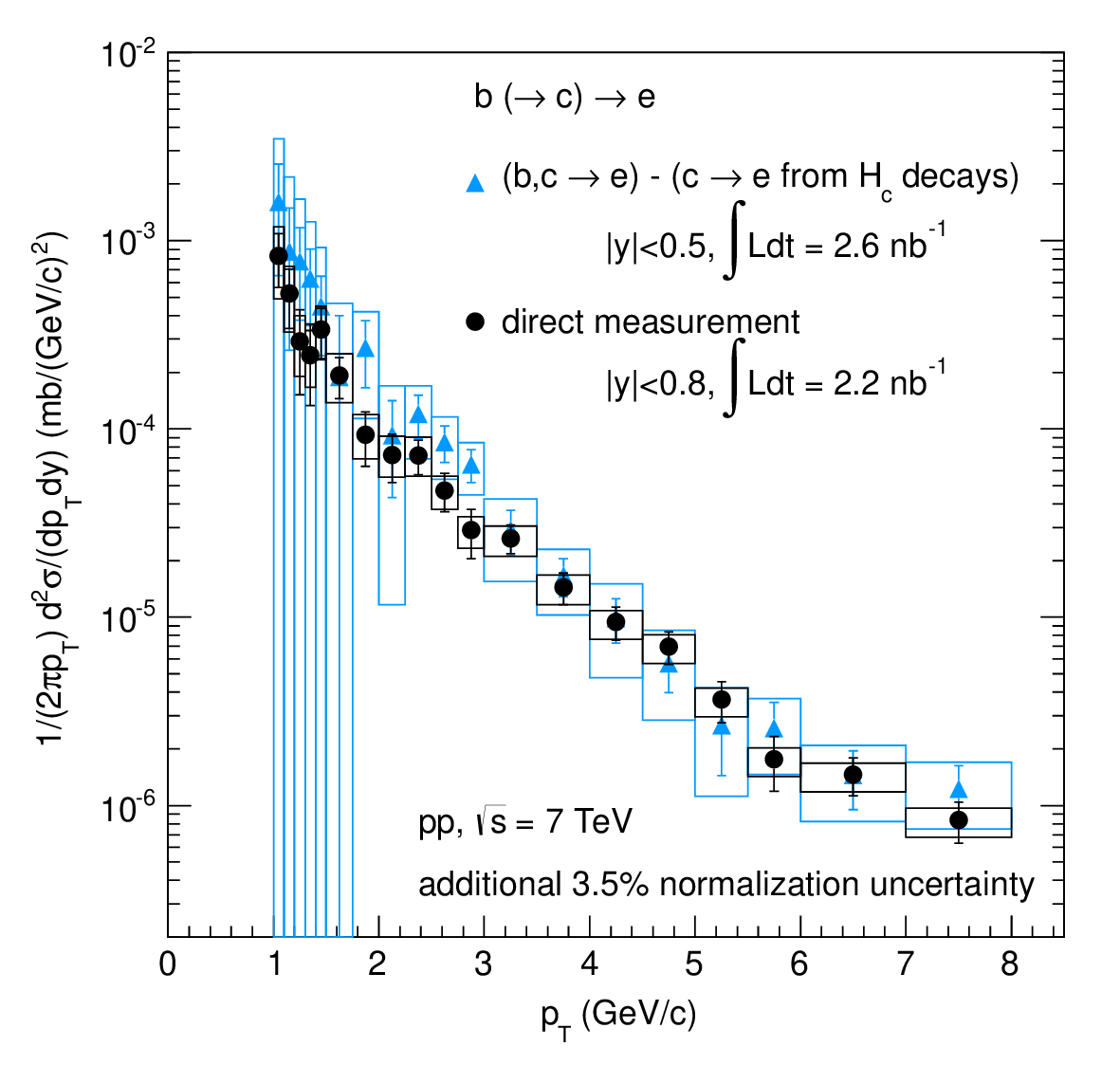}
\caption{\label{fig:fig4}
         (Color online) Invariant cross sections of electrons from beauty 
         hadron decays measured directly via the transverse impact parameter 
         method and indirectly via subtracting the calculated charm hadron 
         decay contribution from the measured heavy-flavor hadron decay 
         electron spectrum~\cite{alice_e}. The error bars (boxes) represent 
         the statistical (systematic) uncertainties.}   
\end{figure}

Figure~\ref{fig:fig4} presents the invariant production cross section of 
electrons from beauty hadron decays obtained with the analysis based on the \dzero cut. 
As a cross check the corresponding result from an alternative
method is shown. In the latter, the decay electron spectrum was calculated for 
charm hadrons as measured with ALICE~\cite{alice_d2h} based on a fast Monte Carlo
simulation using PYTHIA decay kinematics, and it was subtracted 
from the electron spectrum measured for all heavy-flavor hadron 
decays~\cite{alice_e}. The systematic uncertainties of these two inputs have been added 
in quadrature as they are uncorrelated. 
The results from the subtraction method, which does not use a \dzero cut, 
and from the analysis based on the \dzero selection agree within 
the experimental uncertainties, which are much smaller, in particular 
at low \pt, for the beauty measurement employing the \dzero cut. 

\begin{figure}[t]
\centering
\includegraphics[width=0.7\linewidth]{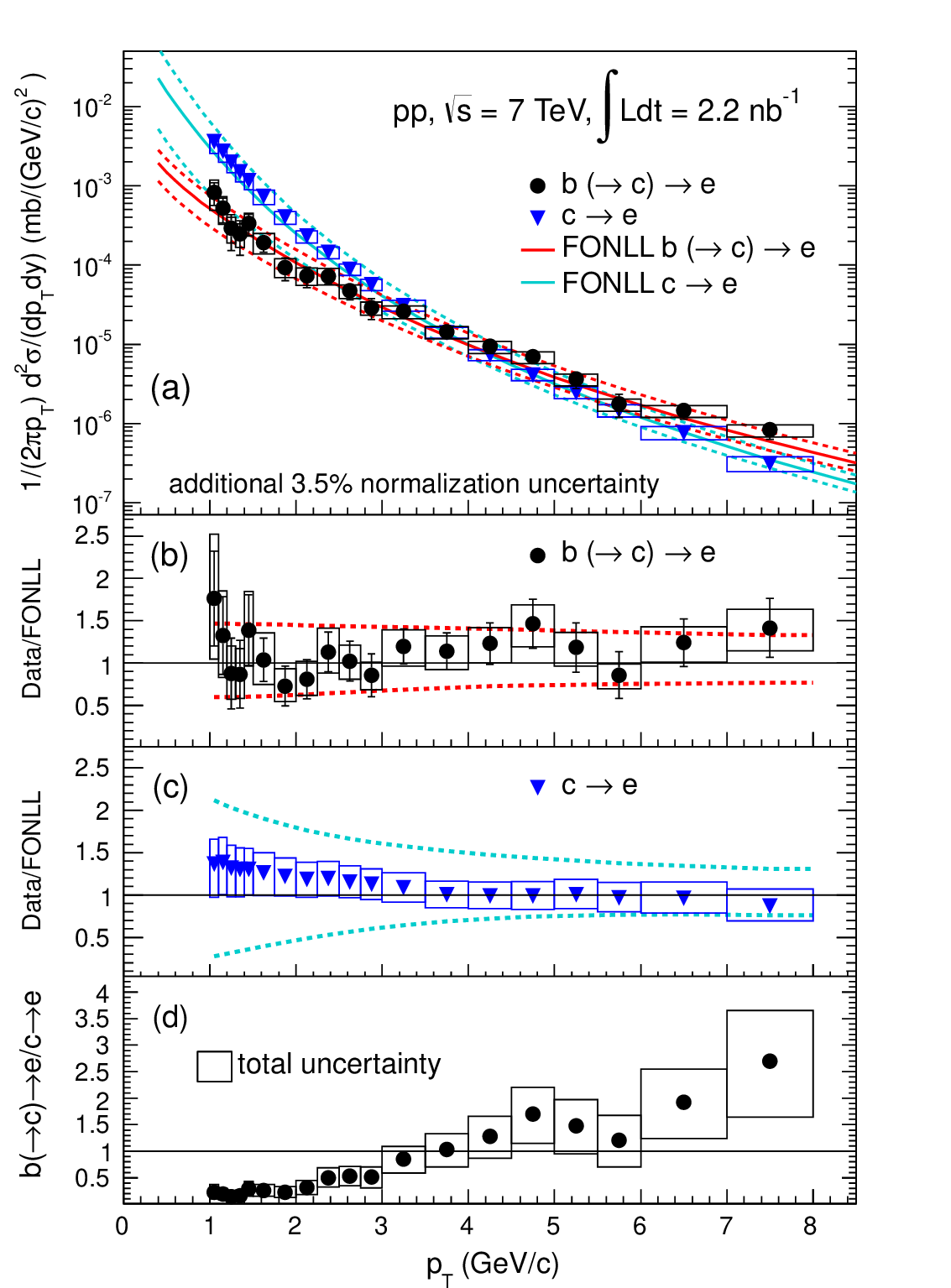}
\caption{\label{fig:fig5}
         (Color online) (a) \pt-differential invariant cross sections of 
         electrons from beauty and from charm hadron decays. The error bars 
         (boxes) represent the statistical (systematic) uncertainties. The
         solid (dashed) lines indicate the corresponding FONLL 
         predictions (uncertainties)~\cite{fonll3}. Ratios of the data and the 
         FONLL calculations are shown in (b) and (c) for electrons from beauty 
         and charm hadron decays, respectively, where the dashed lines indicate
         the FONLL uncertainties. (d) Measured ratio of electrons from beauty 
         and charm hadron decays with error boxes depicting the total 
         uncertainty.}   
\end{figure}

In Fig.~\ref{fig:fig5}(a) FONLL pQCD predictions~\cite{fonll3} of the electron 
production cross sections are compared with the measured electron spectrum 
from beauty hadron decays and with the calculated electron spectrum from charm 
hadron decays. The ratios
of the measured cross sections to the FONLL predictions are shown in 
Fig.~\ref{fig:fig5}(b) and \ref{fig:fig5}(c) for electrons from beauty
and charm hadron decays, respectively. The FONLL predictions are in good 
agreement with the data. At low \pt, electrons from heavy-flavor hadron decays 
originate predominantly from charm hadrons. As demonstrated in 
Fig.~\ref{fig:fig5}(d), beauty hadron decays take over from charm as the 
dominant source of electrons from heavy-flavor hadron decays close to electron 
transverse momenta of 4~\gevc. 
 
The integrated cross section of electrons from beauty hadron decays was measured as 
$6.61 \pm 0.54(\rm stat) ^{+1.92}_{-1.86}(\rm sys)$~$\mu$b for $1< \pt < 8$~\gevc 
in the range $|y|<0.8$. 
The beauty production cross section $\sigmabb$ was calculated by extrapolating
this \pt-integrated visible cross section down to \pt = 0 and to the full $y$ range. 
The extrapolation factor was determined based on FONLL as described in~\cite{alice_jpsi_b}, 
using the beauty to electron branching ratio 
${\rm BR}_{{\rm H_b} \rightarrow {\rm e}} + 
{\rm BR}_{{\rm H_b} \rightarrow {\rm H_c} \rightarrow {\rm e}} 
= 0.205 \pm 0.007$~\cite{pdg}.
The related uncertainty was obtained as the quadratic sum of the uncertainties 
from the beauty quark mass, from perturbative scales, and from the CTEQ6.6 
parton distribution functions~\cite{CTEQ66}.
At mid-rapidity the beauty production cross section per unit rapidity 
is $d\sigma_{\bbbar}/dy = 42.3 \pm 3.5(\rm stat) ^{+12.3}_{-11.9}(\rm sys) 
^{+1.1}_{-1.7}(\rm extr)$~$\mu$b, where the additional systematic uncertainty 
due to the extrapolation procedure is quoted separately.
The total cross section was derived as 
$\sigma_{\bbbar} = 280 \pm 23(\rm stat) ^{+81}_{-79}(\rm sys) ^{+7}_{-8}(\rm extr) 
\pm 10(\rm BR)$~$\mu$b, consistent with the result of a previous measurement 
of \jpsi mesons from beauty hadron decays
$\sigma_{\bbbar} = 282 \pm 74(\rm stat) ^{+58}_{-68}(\rm sys) 
^{+8}_{-7}(\rm extr)$~$\mu$b~\cite{alice_jpsi_b}. The weighted average of
the two measurements was calculated based on the procedure described 
in \cite{Lyons:1988rp}. The statistical and systematic uncertainties of
two measurements are largely uncorrelated, but the extrapolation uncertainties using 
the same theoretical model (FONLL) are correlated. The weights, defined using 
the statistical and the uncorrelated systematic uncertainties, and the correlated 
extrapolation uncertainties, are calculated as 0.499 for the measurement using 
semileptonic beauty hadron decays and 0.501 for that using non-prompt \jpsi mesons. 
The combined total cross section is $\sigma_{\bbbar} = 281 \pm 34(\rm stat) 
^{+53}_{-54}(\rm sys) ^{+7}_{-8}(\rm extr)$~$\mu$b. FONLL predicts
$\sigma_{\bbbar} = 259^{+120}_{-96}$~$\mu$b~\cite{fonll3}.

The production cross section of electrons from heavy-flavor hadron decays 
was measured as $37.7 \pm 3.2(\rm stat) ^{+13.3}_{-14.4}(\rm sys)$~$\mu$b
for $0.5 < \pt < 8$~\gevc in the range $|y| < 0.5$~\cite{alice_e}. 
After subtraction of the contribution from beauty hadron decays (see above) 
the resulting production cross section of electrons from charm hadron decays 
was converted into a charm production cross section applying the same 
extrapolation method as for beauty. With the branching ratio 
${\rm BR}_{{\rm H_c} \rightarrow {\rm e}} = 0.096 \pm 0.004$~\cite{pdg}, 
at mid-rapidity the charm production cross section per unit rapidity is
$d\sigma_{\ccbar}/dy = 1.2 \pm 0.2(\rm stat) \pm 0.6(\rm sys) 
^{+0.2}_{-0.1}(\rm extr)$~mb.
The total cross section $\sigma_{\ccbar} = 10.0 \pm 1.7(\rm stat) 
^{+5.1}_{-5.5}(\rm sys) ^{+3.5}_{-0.5}(\rm extr) \pm 0.4(\rm BR)$~mb 
is consistent with the result of a previous, more accurate measurement using 
D mesons 
$\sigma_{\ccbar} = 8.5 \pm 0.5(\rm stat) ^{+1.0}_{-2.4}(\rm sys) 
^{+5.0}_{-0.4}(\rm extr)$~mb~\cite{alice_charm_xsec}. The FONLL prediction is
$\sigma_{\ccbar} = 4.76^{+6.44}_{-3.25}$~mb~\cite{fonll3}.
All measured cross sections have an additional normalization uncertainty 
of 3.5\%~\cite{alice_cross_section}.
 
In summary, invariant production cross sections of electrons from beauty and 
from charm hadron decays were measured in \pp collisions at $\sqrt{s} = 7$~TeV.
The agreement between theoretical predictions and the data suggests that FONLL 
pQCD calculations can reliably describe heavy-flavor production even at low \pt
in the highest energy hadron collisions accessible in the laboratory today. 
Furthermore, these results provide a crucial baseline for heavy-flavor 
production studies in the hot and dense matter created in \pbpb collisions at 
the LHC.

The ALICE collaboration would like to thank all its engineers and technicians for their invaluable contributions to the construction of the experiment and the CERN accelerator teams for the outstanding performance of the LHC complex.
The ALICE collaboration would like to thank M. Cacciari for providing the FONLL pQCD predictions for the cross sections of
electrons from heavy-flavour hadron decays. 
The ALICE collaboration acknowledges the following funding agencies for their support in building and
running the ALICE detector:
State Committee of Science, Calouste Gulbenkian Foundation from
Lisbon and Swiss Fonds Kidagan, Armenia;
Conselho Nacional de Desenvolvimento Cient\'{\i}fico e Tecnol\'{o}gico (CNPq), Financiadora de Estudos e Projetos (FINEP),
Funda\c{c}\~{a}o de Amparo \`{a} Pesquisa do Estado de S\~{a}o Paulo (FAPESP);
National Natural Science Foundation of China (NSFC), the Chinese Ministry of Education (CMOE)
and the Ministry of Science and Technology of China (MSTC);
Ministry of Education and Youth of the Czech Republic;
Danish Natural Science Research Council, the Carlsberg Foundation and the Danish National Research Foundation;
The European Research Council under the European Community's Seventh Framework Programme;
Helsinki Institute of Physics and the Academy of Finland;
French CNRS-IN2P3, the `Region Pays de Loire', `Region Alsace', `Region Auvergne' and CEA, France;
German BMBF and the Helmholtz Association;
General Secretariat for Research and Technology, Ministry of
Development, Greece;
Hungarian OTKA and National Office for Research and Technology (NKTH);
Department of Atomic Energy and Department of Science and Technology of the Government of India;
Istituto Nazionale di Fisica Nucleare (INFN) and Centro Fermi -
Museo Storico della Fisica e Centro Studi e Ricerche "Enrico
Fermi", Italy;
MEXT Grant-in-Aid for Specially Promoted Research, Ja\-pan;
Joint Institute for Nuclear Research, Dubna;
National Research Foundation of Korea (NRF);
CONACYT, DGAPA, M\'{e}xico, ALFA-EC and the HELEN Program (High-Energy physics Latin-American--European Network);
Stichting voor Fundamenteel Onderzoek der Materie (FOM) and the Nederlandse Organisatie voor Wetenschappelijk Onderzoek (NWO), Netherlands;
Research Council of Norway (NFR);
Polish Ministry of Science and Higher Education;
National Authority for Scientific Research - NASR (Autoritatea Na\c{t}ional\u{a} pentru Cercetare \c{S}tiin\c{t}ific\u{a} - ANCS);
Ministry of Education and Science of Russian Federation,
International Science and Technology Center, Russian Academy of
Sciences, Russian Federal Agency of Atomic Energy, Russian Federal
Agency for Science and Innovations and CERN-INTAS;
Ministry of Education of Slovakia;
Department of Science and Technology, South Africa;
CIEMAT, EELA, Ministerio de Educaci\'{o}n y Ciencia of Spain, Xunta de Galicia (Conseller\'{\i}a de Educaci\'{o}n),
CEA\-DEN, Cubaenerg\'{\i}a, Cuba, and IAEA (International Atomic Energy Agency);
Swedish Research Council (VR) and Knut $\&$ Alice Wallenberg
Foundation (KAW);
Ukraine Ministry of Education and Science;
United Kingdom Science and Technology Facilities Council (STFC);
The United States Department of Energy, the United States National
Science Foundation, the State of Texas, and the State of Ohio.

\bibliography{hfe_pp_beauty}   
\newpage
\begin{appendix}
\section{The ALICE Collaboration}
\label{authorlist}

\begingroup
\small
\begin{flushleft}
B.~Abelev\Irefn{org1234}\And
J.~Adam\Irefn{org1274}\And
D.~Adamov\'{a}\Irefn{org1283}\And
A.M.~Adare\Irefn{org1260}\And
M.M.~Aggarwal\Irefn{org1157}\And
G.~Aglieri~Rinella\Irefn{org1192}\And
A.G.~Agocs\Irefn{org1143}\And
A.~Agostinelli\Irefn{org1132}\And
S.~Aguilar~Salazar\Irefn{org1247}\And
Z.~Ahammed\Irefn{org1225}
% ...
\renewcommand\labelenumi{\textsuperscript{\theenumi}~}
\section*{Affiliation notes}
\renewcommand\theenumi{\roman{enumi}}
\begin{Authlist}
\item \Adef{M.V.Lomonosov Moscow State University, D.V.Skobeltsyn Institute of Nuclear Physics, Moscow, Russia}Also at: M.V.Lomonosov Moscow State University, D.V.Skobeltsyn Institute of Nuclear Physics, Moscow, Russia
\item \Adef{University of Belgrade, Faculty of Physics and Institute of Nuclear Sciences, Belgrade, Serbia}Also at: University of Belgrade, Faculty of Physics and "Vin\v{c}a" Institute of Nuclear Sciences, Belgrade, Serbia
\end{Authlist}
\section*{Collaboration Institutes}
\renewcommand\theenumi{\arabic{enumi}~}
\begin{Authlist}
\item \Idef{org1279}Benem\'{e}rita Universidad Aut\'{o}noma de Puebla, Puebla, Mexico
\item \Idef{org1220}Bogolyubov Institute for Theoretical Physics, Kiev, Ukraine
\item \Idef{org1262}Budker Institute for Nuclear Physics, Novosibirsk, Russia
\item \Idef{org1292}California Polytechnic State University, San Luis Obispo, California, United States
\item \Idef{org1329}Central China Normal University, Wuhan, China
\item \Idef{org14939}Centre de Calcul de l'IN2P3, Villeurbanne, France
\item \Idef{org1197}Centro de Aplicaciones Tecnol\'{o}gicas y Desarrollo Nuclear (CEADEN), Havana, Cuba
\item \Idef{org1242}Centro de Investigaciones Energ\'{e}ticas Medioambientales y Tecnol\'{o}gicas (CIEMAT), Madrid, Spain
\item \Idef{org1244}Centro de Investigaci\'{o}n y de Estudios Avanzados (CINVESTAV), Mexico City and M\'{e}rida, Mexico
\item \Idef{org1335}Centro Fermi -- Centro Studi e Ricerche e Museo Storico della Fisica ``Enrico Fermi'', Rome, Italy
\item \Idef{org17347}Chicago State University, Chicago, United States
\item \Idef{org1288}Commissariat \`{a} l'Energie Atomique, IRFU, Saclay, France
\item \Idef{org1294}Departamento de F\'{\i}sica de Part\'{\i}culas and IGFAE, Universidad de Santiago de Compostela, Santiago de Compostela, Spain
\item \Idef{org1106}Department of Physics Aligarh Muslim University, Aligarh, India
\item \Idef{org1121}Department of Physics and Technology, University of Bergen, Bergen, Norway
\item \Idef{org1162}Department of Physics, Ohio State University, Columbus, Ohio, United States
\item \Idef{org1300}Department of Physics, Sejong University, Seoul, South Korea
\item \Idef{org1268}Department of Physics, University of Oslo, Oslo, Norway
\item \Idef{org1132}Dipartimento di Fisica dell'Universit\`{a} and Sezione INFN, Bologna, Italy
\item \Idef{org1270}Dipartimento di Fisica dell'Universit\`{a} and Sezione INFN, Padova, Italy
\item \Idef{org1315}Dipartimento di Fisica dell'Universit\`{a} and Sezione INFN, Trieste, Italy
\item \Idef{org1145}Dipartimento di Fisica dell'Universit\`{a} and Sezione INFN, Cagliari, Italy
\item \Idef{org1312}Dipartimento di Fisica dell'Universit\`{a} and Sezione INFN, Turin, Italy
\item \Idef{org1285}Dipartimento di Fisica dell'Universit\`{a} `La Sapienza' and Sezione INFN, Rome, Italy
\item \Idef{org1154}Dipartimento di Fisica e Astronomia dell'Universit\`{a} and Sezione INFN, Catania, Italy
\item \Idef{org1290}Dipartimento di Fisica `E.R.~Caianiello' dell'Universit\`{a} and Gruppo Collegato INFN, Salerno, Italy
\item \Idef{org1103}Dipartimento di Scienze e Innovazione Tecnologica dell'Universit\`{a} del Piemonte Orientale and Gruppo Collegato INFN, Alessandria, Italy
\item \Idef{org1114}Dipartimento Interateneo di Fisica `M.~Merlin' and Sezione INFN, Bari, Italy
\item \Idef{org1237}Division of Experimental High Energy Physics, University of Lund, Lund, Sweden
\item \Idef{org1192}European Organization for Nuclear Research (CERN), Geneva, Switzerland
\item \Idef{org1227}Fachhochschule K\"{o}ln, K\"{o}ln, Germany
\item \Idef{org1122}Faculty of Engineering, Bergen University College, Bergen, Norway
\item \Idef{org1136}Faculty of Mathematics, Physics and Informatics, Comenius University, Bratislava, Slovakia
\item \Idef{org1274}Faculty of Nuclear Sciences and Physical Engineering, Czech Technical University in Prague, Prague, Czech Republic
\item \Idef{org1229}Faculty of Science, P.J.~\v{S}af\'{a}rik University, Ko\v{s}ice, Slovakia
\item \Idef{org1184}Frankfurt Institute for Advanced Studies, Johann Wolfgang Goethe-Universit\"{a}t Frankfurt, Frankfurt, Germany
\item \Idef{org1215}Gangneung-Wonju National University, Gangneung, South Korea
\item \Idef{org1212}Helsinki Institute of Physics (HIP) and University of Jyv\"{a}skyl\"{a}, Jyv\"{a}skyl\"{a}, Finland
\item \Idef{org1203}Hiroshima University, Hiroshima, Japan
\item \Idef{org1254}Indian Institute of Technology Bombay (IIT), Mumbai, India
\item \Idef{org36378}Indian Institute of Technology Indore (IIT), Indore, India
\item \Idef{org1266}Institut de Physique Nucl\'{e}aire d'Orsay (IPNO), Universit\'{e} Paris-Sud, CNRS-IN2P3, Orsay, France
\item \Idef{org1277}Institute for High Energy Physics, Protvino, Russia
\item \Idef{org1249}Institute for Nuclear Research, Academy of Sciences, Moscow, Russia
\item \Idef{org1320}Nikhef, National Institute for Subatomic Physics and Institute for Subatomic Physics of Utrecht University, Utrecht, Netherlands
\item \Idef{org1250}Institute for Theoretical and Experimental Physics, Moscow, Russia
\item \Idef{org1230}Institute of Experimental Physics, Slovak Academy of Sciences, Ko\v{s}ice, Slovakia
\item \Idef{org1127}Institute of Physics, Bhubaneswar, India
\item \Idef{org1275}Institute of Physics, Academy of Sciences of the Czech Republic, Prague, Czech Republic
\item \Idef{org1139}Institute of Space Sciences (ISS), Bucharest, Romania
\item \Idef{org27399}Institut f\"{u}r Informatik, Johann Wolfgang Goethe-Universit\"{a}t Frankfurt, Frankfurt, Germany
\item \Idef{org1185}Institut f\"{u}r Kernphysik, Johann Wolfgang Goethe-Universit\"{a}t Frankfurt, Frankfurt, Germany
\item \Idef{org1177}Institut f\"{u}r Kernphysik, Technische Universit\"{a}t Darmstadt, Darmstadt, Germany
\item \Idef{org1256}Institut f\"{u}r Kernphysik, Westf\"{a}lische Wilhelms-Universit\"{a}t M\"{u}nster, M\"{u}nster, Germany
\item \Idef{org1246}Instituto de Ciencias Nucleares, Universidad Nacional Aut\'{o}noma de M\'{e}xico, Mexico City, Mexico
\item \Idef{org1247}Instituto de F\'{\i}sica, Universidad Nacional Aut\'{o}noma de M\'{e}xico, Mexico City, Mexico
\item \Idef{org23333}Institut of Theoretical Physics, University of Wroclaw
\item \Idef{org1308}Institut Pluridisciplinaire Hubert Curien (IPHC), Universit\'{e} de Strasbourg, CNRS-IN2P3, Strasbourg, France
\item \Idef{org1182}Joint Institute for Nuclear Research (JINR), Dubna, Russia
\item \Idef{org1143}KFKI Research Institute for Particle and Nuclear Physics, Hungarian Academy of Sciences, Budapest, Hungary
\item \Idef{org1199}Kirchhoff-Institut f\"{u}r Physik, Ruprecht-Karls-Universit\"{a}t Heidelberg, Heidelberg, Germany
\item \Idef{org20954}Korea Institute of Science and Technology Information, Daejeon, South Korea
\item \Idef{org1160}Laboratoire de Physique Corpusculaire (LPC), Clermont Universit\'{e}, Universit\'{e} Blaise Pascal, CNRS--IN2P3, Clermont-Ferrand, France
\item \Idef{org1194}Laboratoire de Physique Subatomique et de Cosmologie (LPSC), Universit\'{e} Joseph Fourier, CNRS-IN2P3, Institut Polytechnique de Grenoble, Grenoble, France
\item \Idef{org1187}Laboratori Nazionali di Frascati, INFN, Frascati, Italy
\item \Idef{org1232}Laboratori Nazionali di Legnaro, INFN, Legnaro, Italy
\item \Idef{org1125}Lawrence Berkeley National Laboratory, Berkeley, California, United States
\item \Idef{org1234}Lawrence Livermore National Laboratory, Livermore, California, United States
\item \Idef{org1251}Moscow Engineering Physics Institute, Moscow, Russia
\item \Idef{org1140}National Institute for Physics and Nuclear Engineering, Bucharest, Romania
\item \Idef{org1165}Niels Bohr Institute, University of Copenhagen, Copenhagen, Denmark
\item \Idef{org1109}Nikhef, National Institute for Subatomic Physics, Amsterdam, Netherlands
\item \Idef{org1283}Nuclear Physics Institute, Academy of Sciences of the Czech Republic, \v{R}e\v{z} u Prahy, Czech Republic
\item \Idef{org1264}Oak Ridge National Laboratory, Oak Ridge, Tennessee, United States
\item \Idef{org1189}Petersburg Nuclear Physics Institute, Gatchina, Russia
\item \Idef{org1170}Physics Department, Creighton University, Omaha, Nebraska, United States
\item \Idef{org1157}Physics Department, Panjab University, Chandigarh, India
\item \Idef{org1112}Physics Department, University of Athens, Athens, Greece
\item \Idef{org1152}Physics Department, University of Cape Town, iThemba LABS, Cape Town, South Africa
\item \Idef{org1209}Physics Department, University of Jammu, Jammu, India
\item \Idef{org1207}Physics Department, University of Rajasthan, Jaipur, India
\item \Idef{org1200}Physikalisches Institut, Ruprecht-Karls-Universit\"{a}t Heidelberg, Heidelberg, Germany
\item \Idef{org1325}Purdue University, West Lafayette, Indiana, United States
\item \Idef{org1281}Pusan National University, Pusan, South Korea
\item \Idef{org1176}Research Division and ExtreMe Matter Institute EMMI, GSI Helmholtzzentrum f\"ur Schwerionenforschung, Darmstadt, Germany
\item \Idef{org1334}Rudjer Bo\v{s}kovi\'{c} Institute, Zagreb, Croatia
\item \Idef{org1298}Russian Federal Nuclear Center (VNIIEF), Sarov, Russia
\item \Idef{org1252}Russian Research Centre Kurchatov Institute, Moscow, Russia
\item \Idef{org1224}Saha Institute of Nuclear Physics, Kolkata, India
\item \Idef{org1130}School of Physics and Astronomy, University of Birmingham, Birmingham, United Kingdom
\item \Idef{org1338}Secci\'{o}n F\'{\i}sica, Departamento de Ciencias, Pontificia Universidad Cat\'{o}lica del Per\'{u}, Lima, Peru
\item \Idef{org1316}Sezione INFN, Trieste, Italy
\item \Idef{org1271}Sezione INFN, Padova, Italy
\item \Idef{org1313}Sezione INFN, Turin, Italy
\item \Idef{org1286}Sezione INFN, Rome, Italy
\item \Idef{org1146}Sezione INFN, Cagliari, Italy
\item \Idef{org1133}Sezione INFN, Bologna, Italy
\item \Idef{org1115}Sezione INFN, Bari, Italy
\item \Idef{org1155}Sezione INFN, Catania, Italy
\item \Idef{org1322}Soltan Institute for Nuclear Studies, Warsaw, Poland
\item \Idef{org36377}Nuclear Physics Group, STFC Daresbury Laboratory, Daresbury, United Kingdom
\item \Idef{org1258}SUBATECH, Ecole des Mines de Nantes, Universit\'{e} de Nantes, CNRS-IN2P3, Nantes, France
\item \Idef{org1304}Technical University of Split FESB, Split, Croatia
\item \Idef{org1168}The Henryk Niewodniczanski Institute of Nuclear Physics, Polish Academy of Sciences, Cracow, Poland
\item \Idef{org17361}The University of Texas at Austin, Physics Department, Austin, TX, United States
\item \Idef{org1173}Universidad Aut\'{o}noma de Sinaloa, Culiac\'{a}n, Mexico
\item \Idef{org1296}Universidade de S\~{a}o Paulo (USP), S\~{a}o Paulo, Brazil
\item \Idef{org1149}Universidade Estadual de Campinas (UNICAMP), Campinas, Brazil
\item \Idef{org1239}Universit\'{e} de Lyon, Universit\'{e} Lyon 1, CNRS/IN2P3, IPN-Lyon, Villeurbanne, France
\item \Idef{org1205}University of Houston, Houston, Texas, United States
\item \Idef{org20371}University of Technology and Austrian Academy of Sciences, Vienna, Austria
\item \Idef{org1222}University of Tennessee, Knoxville, Tennessee, United States
\item \Idef{org1310}University of Tokyo, Tokyo, Japan
\item \Idef{org1318}University of Tsukuba, Tsukuba, Japan
\item \Idef{org21360}Eberhard Karls Universit\"{a}t T\"{u}bingen, T\"{u}bingen, Germany
\item \Idef{org1225}Variable Energy Cyclotron Centre, Kolkata, India
\item \Idef{org1306}V.~Fock Institute for Physics, St. Petersburg State University, St. Petersburg, Russia
\item \Idef{org1323}Warsaw University of Technology, Warsaw, Poland
\item \Idef{org1179}Wayne State University, Detroit, Michigan, United States
\item \Idef{org1260}Yale University, New Haven, Connecticut, United States
\item \Idef{org1332}Yerevan Physics Institute, Yerevan, Armenia
\item \Idef{org15649}Yildiz Technical University, Istanbul, Turkey
\item \Idef{org1301}Yonsei University, Seoul, South Korea
\item \Idef{org1327}Zentrum f\"{u}r Technologietransfer und Telekommunikation (ZTT), Fachhochschule Worms, Worms, Germany
\end{Authlist}
\end{flushleft}
\endgroup

\end{appendix}

\end{document}